\documentclass[useAMS,usenatbib]{mn2e}

\usepackage[dvips]{graphicx}
\usepackage{float}
\usepackage[dvips]{color}
\usepackage{hyphenat}
\usepackage{float}
\usepackage{subfigure}
\begin{document}

\newcommand{\NII}{{[\sc N\,i\,i]}}
\newcommand{\HII}{{\sc H\,ii}}
\newcommand{\Ha}{H$\alpha$}
\newcommand{\Haw}{\hbox{H$\alpha$\,$\lambda $6562.78}}
\newcommand{\Hbw}{\hbox{H$\beta$\,$\lambda $4861}}
\newcommand{\ghafas}{{\sc GH$\alpha$FaS}}

\def\kms{$\mbox{km s}^{-1}$}
\def\Myr{$\mbox{M}_\odot\mbox{ yr}^{-1}$}
\def\etal{et al.~}
\def\deg{^\circ}
\def\asim{\mathord{\sim}}
\def\farcs{\hbox{$.\!\!^{\prime\prime}$}}
\def\ergs{$erg\, s^{-1}$}

\def\javi{\color{blue}}
\def\kambiz{\color{magenta}}
\def\comment{\color{green}}
\def\black{\color{black}}


\title[Statistical studies of the internal kinematics of \HII\  regions: the case of M\,83]{An improved method for statistical studies of the internal kinematics of \HII\ regions: the case of M\,83}

\author[Blasco-Herrera et al.]{Blasco-Herrera, J.$^{1}$\thanks{E-mail: javier@astro.su.se}, 
Fathi, K.$^{1,2}$, 
Beckman J.$^{3,4,5}$, 
Guti\'errez L.$^{3,6}$, 
Lundgren A.$^{7}$, \newauthor 
Epinat, B.$^{8}$, 
\"Ostlin, G.$^{1,2}$, 
Font, J.$^{3}$, 
Hernandez O.$^{9}$,
de Denus-Baillargeon, M-M.$^{9,10}$, \newauthor
Carignan, C.$^{9,11}$,  \\ \ \\
$^1$Stockholm Observatory, Department of Astronomy, Stockholm University, AlbaNova Center, 106 91 Stockholm, Sweden\\
$^2$Oskar Klein Centre for Cosmoparticle Physics, Stockholm University, 106 91 Stockholm, Sweden\\
$^3$Instituto de Astrof\'\i sica de Canarias, C/ V\'\i a L\'actea s/n, 38200 La Laguna, Tenerife, Spain\\
$^4$Consejo Superior de Investigaciones Cient\'\i ficas, Spain\\
$^5$Departamento de Astrofísica, Universidad de La Laguna, Tenerife, Spain\\
$^6$Universidad Nacional Aut\'onoma de M\'exico, Apartado Postal 877, Ensenada, B. C. 22800, M\'exico\\
$^{7}$European Southern Observatory, Casilla 19001, Santiago 19, Chile \\
$^{8}$Laboratoire d'Astrophysique de Toulouse-Tarbes, Universit\'e de Toulouse, CNRS, 14 Avenue \'Edouard Belin, 31400 Toulouse, France\\
$^9$Laboratoire d'Astrophysique Exp\'erimentale , Universit\'e de Montr\'eal, C.P. 6128 succ. centre ville, Montr\'eal, QC, Canada H3C 3J7\\
$^{10}$Institut Fresnel, CNRS, Univ. Aix Marseille, France\\
$^{11}$Observatoire d'Astrophysique de l'Université de Ouagadougou (UFR/SEA), 03 BP 7021 Ouagadougou 03, Burkina Faso}

\date{Accepted for publication, 2010 May 24}
\pagerange{\pageref{firstpage}--\pageref{lastpage}} \pubyear{2008}
\maketitle

\label{firstpage}

\begin{abstract}
We present the integrated \Ha\ emission line profile for 157 \HII\ regions in the central $3.4'\times 3.4'$ of the galaxy M\,83 (NGC\,5236). Using the Fabry-Perot interferometer \ghafas, on the 4.2 m William Herschel on La Palma, we show the importance of a good characterization of the instrumental response function for the study of line profile shapes. 

The luminosity-velocity dispersion relation is also studied, and in the $\log(L)-\log(\sigma)$ plane we do not find a linear relation, but an upper envelope with equation $\log(L_{{\rm H}\alpha})=0.9 \cdot \log(\sigma)+38.1$. For the adopted distance of 4.5 Mpc, the upper envelope appears at the luminosity $L=10^{ 38.5}$ \ergs, in full agreement with previous studies of other galaxies, reinforcing the idea of using \HII\ regions as standard candles. 
\end{abstract}


\section{Introduction}
\label{sec:intro}

\HII\ regions are among the brightest non-ephemeral objects in galaxies, with prominent emission lines characterizing their spectra. By studying the integrated line profiles of such emission lines we can, to some extent, infer the physical properties inside the regions, the virial mass of the system, temperatures, densities, and their ages. One important parameter for deriving dynamical properties is the velocity dispersion of the emission line $\sigma={\rm FWHM}/2\sqrt{2\ln 2}$, where FWHM is the full width at half maximum of the line.
By studying the integrated \Ha\ emission-line profiles of giant extra-galactic \HII\ regions, \cite{SmithandWeedman70} found velocity dispersions, $\sigma \sim 25$ \kms. Since then, great efforts have been made to explain the line profiles of \HII\ regions. \cite{Melnick77} and \cite{TerlevichandMelnick81} analyzed 12 regions and claimed to find a relation between sizes, luminosities $L$, and integrated emission line velocity dispersion $\sigma$ \citep[for further discussion in this topic see][]{{Melnicketal88},{TellesandTerlevich93}}. One fundamental relation reported by these authors is $L \propto \sigma^4$, which is equivalent to the Faber-Jackson relation for elliptical galaxies \citep{FaberandJackson76}. Given the potential importance of finding a valid direct relation between $L$ and $\sigma$ to understand the internal dynamics of \HII\ regions but that observations have not seemed to yield a unique relation, it has been the subject of some controversy.
 
After an initial claim by \cite{TerlevichandMelnick81} that they had observed a relation $L \propto \sigma^4$ implying that \HII\ regions are in virial equilibrium,

\cite{Royetal86} and \cite{Hippelein86} found a similar relation, but with exponents $\sim 3$ and $\sim 6$, respectively, while \cite{GallagherandHunter83} found no clear proportionality between $L$ and $\sigma$. On the other hand, \cite{Melnicketal87} adduced problems in the selection of the sample made by Gallagher \& Hunter, since they used regions with low velocity dispersion, while \cite{TerlevichandMelnick81} used giant \HII\ regions, with high velocity dispersion ($\sigma >13$ \kms ) \black. \cite{Melnicketal87} reported a new value of $\sim 5$ for the exponent in the $L-\sigma$ relation. A common characteristic of all these studies is that they use at most few tens of the brightest regions, most of them in the high velocity dispersion range \black, taken from different galaxies, i.e., introducing extra uncertainties from the assumed distances to different systems. This produces additional scatter as extinction and distances are different for each galaxy. 

The obvious next step was then to study all the regions available from the same galaxy where the distance would have no effect on the scatter and the extragalactic and Galactic extinction would be similar. \cite{Arsenaultetal90} studied 127 regions in the galaxy NGC\,4321 and found no clear $L-\sigma$ relation, except for the subsample of the highest surface brightness regions, for which they report an exponent of $\sim 2.6$. Probably the most remarkable aspect of the study by \cite{Arsenaultetal90} is that they define an upper envelope for the $\log(L)-\log(\sigma)$ diagram, an upper limit for the luminosity at a certain velocity width. This envelope was also found by \cite{Rozasetal98} and \cite{Relanoetal05} with slopes of $2.6$ and $2.0$ respectively, and again by \cite{Rozasetal06a} who found 2.9 for a sample of supergiant \HII\ regions. All these papers relate the upper envelope to where the regions in virial equilibrium are located, while broader line profiles have contribution from non-equilibrium processes

 The line profiles of spatially unresolved \HII\ regions have been traditionally fitted by Gaussians  \citep[][ and many others]{{SmithandWeedman70},{TerlevichandMelnick81}, {Arsenaultetal90}}. \cite{Rozasetal06b} and \cite{Westmoquetteetal07} distinguish a broad low intensity component added to the main spectral component.

This has been more a question of mathematical simplicity than of the physical processes which influence the line shape. Studies of spatially resolved \HII\ regions \citep[e.g.][]{{MeaburnandWalsh81},{Meaburn84}, {Clayton86}, {ChuandKennicutt94}} showed clearly that not only do they receive multiple inputs of energy, but that these inputs vary from place to place (and time to time) within a region, so that the resulting integrated emission line profile is a luminosity weighted mean of all these inputs. We can pick out two principal contributors to the final line shape: the turbulent motion within the ionized gas (particularly where it interacts with the internal and surrounding neutral gas), and the emission from dominant expanding shells \citep[][]{{Dyson79},{Meaburn80},{GallagherandHunter83}}. The contribution from the turbulent motions is gaussian, as it is produced by random motion, while the expanding shells, if present, are identified as equidistant in velocity from the main peak. These symmetric components produce non-gaussian wings or, if the expansion velocity and the luminosity are high enough, secondary peaks. In addition to these two contributions, there may be more local expansion due to the winds of multiple individual stars and supernova shells \citep{Redmanetal03}. Also the hyperfine structure of Hydrogen split \Ha\ in seven components $\sim$0.14 \AA\  apart, which should be considered.

With such a complex ensemble of contributions it is importantthe to take into account the shape of the spectral instrument response to accurately correct its effect on observations. Although, again for the sake of mathematical simplicity, it has traditionally been assumed Gaussian, the instrumental response is better described by a Lorentzian \citep[e.g.][]{Bland-Hawthorn95}. Thus, instead of fitting a Gaussian (assuming the turbulent motion as the dominant feature in the line profile), \cite{MoiseevandEgorov08} concluded that Voigt profiles are a more realistic characterization of the integrated emission line profiles of \HII\ regions, since it meant assuming the instrumental response is Lorentzian, not Gaussian (more about this will be discussed in sect.~\ref{sec:fitting}).

During the present study we take particular care of the instrumental response and fit Gaussians convolved with it to the integrated \Ha\ line profiles, comparing it with the direct fit to a Gaussian. In sect.~\ref{sec:observations} we describe the observations, sect.~\ref{sec:reduction} explains the data reduction, focusing on the improved steps carried out for the present study. The line profiles are analyzed in sect.~\ref{sec:analysis}, with special emphasis in the $L-\sigma$ relation. Results and discussion are in sect.~\ref{sec:results} and conclusions presented in \ref{sec:conclusions}.

\section{Observations}
\label{sec:observations}

We use a narrow-band \Ha\ image to identify and measure the \Ha\ emission flux from the \HII\ regions in M\,83, and use complementary scanning Fabry-Perot observations over the central $\approx 2.3$ kpc radius to derive the dynamical information from their integrated spectra.

\subsection{\Ha\  imaging}
\label{danish_data}
The narrow-band \Ha\ observations were carried out in April 2003 with the Danish 1.54\,m telescope, equipped with DFOSC, at ESO, La Silla, Chile. On the 2k$\times$2k chip, the pixel scale was 0$\farcs$39, and the field of view was 13$\farcm$7$\times$13$\farcm$7. We used the narrow-band filters ESO \#693 ($\lambda_{\rm eff}$\,$\approx$\,6562.3\,{\AA}, $\Delta \lambda$\,$\approx$\,62.1\,{\AA}) and \#697 ($\lambda_{\rm eff}$\,$\approx$\,6654.4\,{\AA}, $\Delta \lambda$\,$\approx$\,61.6\,{\AA}). The latter was used for continuum subtraction after matching with the \Ha\ image using 17 stars to align the two images. In total the exposure times were 2100\,s and 1500\,s in the two filters, respectively. The data reduction was performed in the usual manner including bias level correction, flat-fielding, background correction, flux calibration, foreground star removal, and final continuum subtraction. All these reduction steps have been performed using various IRAF packages. A correction for a Galactic extinction of 0.18$^{\rm m}$ was applied. The derived global flux agrees with that obtained by \cite{BellandKennicutt01}.

Our H$\alpha$ filter contains the \NII\ lines at 6548 and 6583\,{\AA} at either side. The flux ratio of the stronger \NII\ line (6548 and 6583\,{\AA}) to that of the H$\alpha$ line has been measured to 0.45 in the disk of M\,83 \citep{Boissieretal05}. Considering the transmision curve of filter ESO \#693 we estimate that the \NII\ contribution to the measured fluxes in the H$\alpha$ filter is 25\%, and we have corrected the fluxes by a factor of 0.75 to compensate for this.

\subsection{ \ghafas\ data}
\label{sec:FPdata}
We observed M\,83 with the \ghafas\ Fabry-Perot interferometer \citep{Hernandezetal08} mounted on the Nasmyth focus of the 4.2 m William Herschel Telescope (WHT) on La Palma. The observations were carried out in two different sets, July 6 of 2007 ($\sim$ 60 min) and March 18 of 2009 ($\sim 90$ min). Using the interference order p=765, \ghafas\ was tuned to a mean Finesse of $\sim$17 (at \Ha) which allow us to scan the \Haw\ emission line at 8.2 \kms\ (0.2 \AA) over a free spectral range of 392 \kms\ (8.6 \AA). \ghafas\ was set in low spatial resolution mode ($512 \times 512$ pixels, 0.4\,\arcsec/pixel), integrating 5 seconds for each one of the 48 channels during 38 cycles. The mean seeing was $0.7\arcsec \pm 0.1\arcsec$ during the first run, and undetermined for the second run due to a malfunction of the instrumentation of the WHT. We do however measure the approximate seeing from our final cube image after adding both observations and find a mean value of $2\pm0.2$ \arcsec\ across the field. 

\begin{figure*}
\centering
{\includegraphics[angle=00, width=\linewidth]{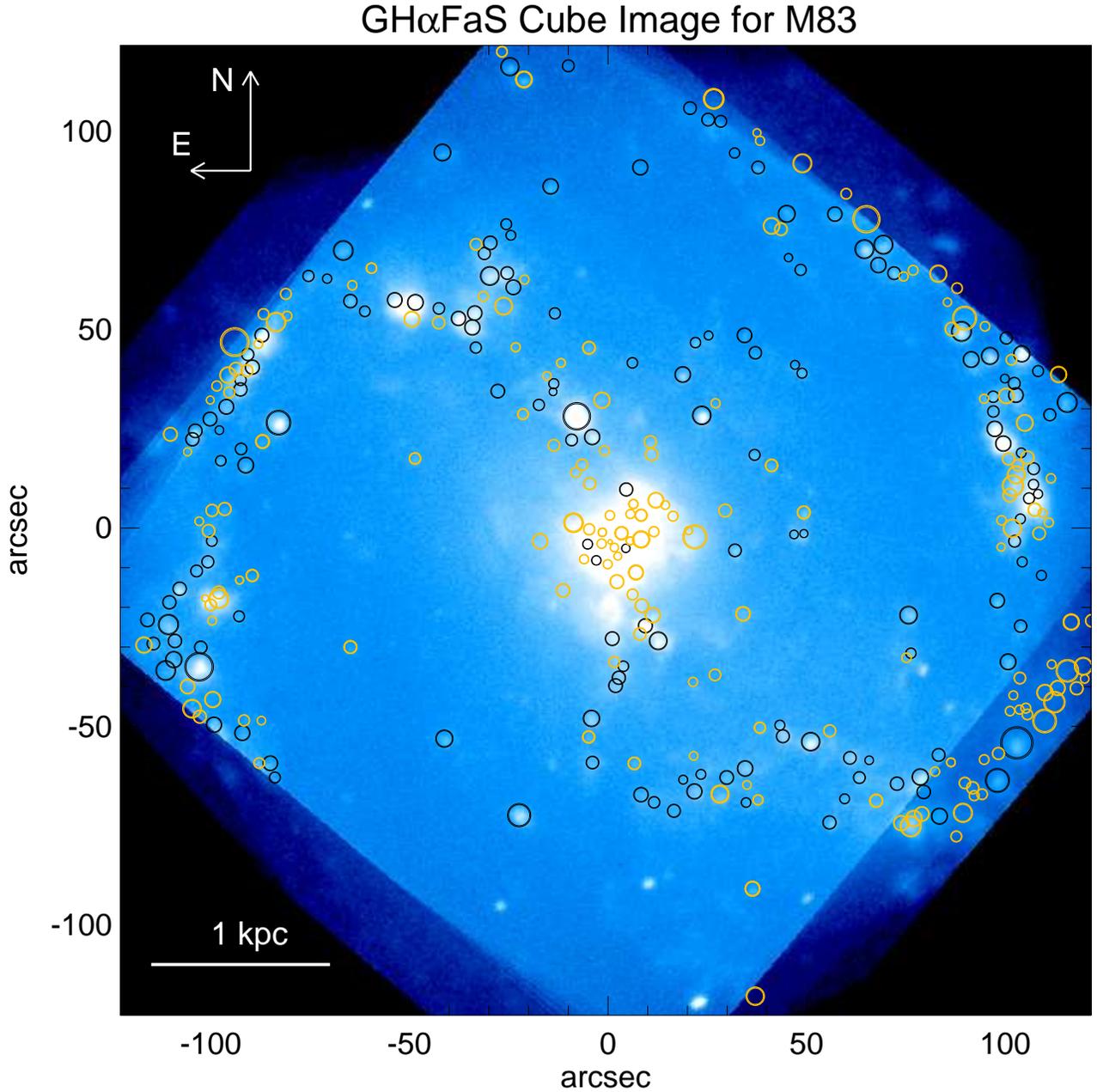}}
\caption{Intensity map of M\,83. Black circles indicate the selected \HII\ regions, yellow circles the rejected ones. The center of the image corresponds to the coordinates RA:13h37m00.8s DEC:-29d51m54s.}\label{fig:selected}
\end{figure*}

The intensity map of the reduced and derotated data cube is illustrated in fig.~\ref{fig:selected}, where all the selected \HII\ regions are marked (see sect.~\ref{sec:leonel} and ~\ref{sec:profiles} for further details).

\section{Data reduction}
\label{sec:reduction}
The commonly adopted data reduction for Fabry-Perot observations aims at deriving line-of-sight velocity fields over extended objects \citep[e.g.][]{{deVaucouleursandPence80}, {Amrametal02}}. The velocity field for M\,83 was presented and analyzed in \cite{Fathietal08} and here we only focus on the shapes of the observed \Ha\ emission line profiles, therefore we need to revise and improve some of the data reduction steps to accommodate the accuracy that is necessary for our analysis. In the following three subsections, we describe the three reduction steps that we have improved: derotation, sky subtraction and instrumental response correction.

\subsection{Derotation}
As \ghafas\ is mounted on the optical table of the Nasmyth focus, the Fabry-Perot observations are affected by the rotation of the field. The optical derotator provided by the Isaac Newton Group of Telescopes has the disadvantage of covering a field of $2.5' \times 2.5'$, which is almost a factor two smaller than the $3.4' \times 3.4'$ field of view of \ghafas. Using a IPCS photomultiplier with rapid readout \citep[][]{{Gachetal02}, {Gachetal03}} we are able to store the image from  individual channels separately. This setup is ideal for derotating the observed cube {\em a posteriori} without any loss of sensitivity or signal. For our observations this means that images with 5 sec interval are stored. We rely on the system speed to record images and improve the procedure developed by \cite{Fathietal08} and \cite{Hernandezetal08} and apply the following scheme:

\begin{enumerate}
 \item Phase-map correcting all individual channels separately: The \emph{phase-map correction} rectifies the position of the incoming photons for the interference rings in each channel, disentangling the wavelength, $\lambda$, from ($x,y$) coordinates and building one sub-cube (with $\lambda$ as the third coordinate) out of each channel.

 \item Stacking channel maps: Given the freedom of packing together any arbitrary number of consecutive channels, we ensure that two criteria are fulfilled: (1) we are able to detect at least three bright point sources in every group of channels, and (2) we see no sign of field rotation within them (ratio between x and y axis for point sources show less than 15\% difference in the stacked images). An image is built for every such group and is analyzed with FIND procedure \citep[part of IDL astrolib library, ][]{Landsman93} until our criteria are fulfilled.

 \item Searching and storing: The coordinates of point sources detected in all the stacked images are calculated by determining the center of the point sources using a centroid.

 \item Calculating the rotation and translation: For every stacked image we ensure that all identified sources coincide with the reference source. We use the POWELL routine of IDL, to compute the transformation $(\Delta x, \Delta y, \Delta \theta)$ that minimizes the maximum deviation of any of the point sources after correcting. Other criteria, such as minimizing the average of the deviations can be easily implemented in the code.

 \item Finally, we apply the calculated rotation and translation to all the sub-cubes and adding them in the final derotated cube.
\end{enumerate}

Re-measuring the position of the sources in all the derotated images, we are able to estimate the errors committed in the process. These errors mainly arise from the marginal uncertainties in the determination of the center of the point sources due to the combination of faint fluxes, smearing of the data by seeing, and the finite spatial sampling element.

It is worth noting that the decision about the integration time is critical for the derotation. We are stacking the channel maps under the assumption that the field rotation can be ignored within the stacked images while at the same time achieving enough signal to accurately measuring the center of the sources. This equilibrium is even more difficult to achieve when some of the point-like sources are emission-line objects, as they will appear in a limited number of channels. Hence, if the reference points are mainly continuum sources, short exposure times are better, while if the derotation relies mainly in emission-line sources longer exposure times are preferred, so that the center of such objects will be well measured.   

In the case of M\,83 we use 14 \HII\ regions and field stars evenly distributed throughout the field and 5 second exposure time per channel. We estimate the accuracy of the derotation to be  $\sim 1$ pixels, with limit cases of $\sim 1.5$ from comparing the position of the sources in the derotated images before adding them in the final cube. Our analysis deals with integrated line profiles, co-adding 22-600 pixels, so we find the errors induced by the derotation negligible.

\subsection{Sky subtraction}
   \label{skysub}

\cite{Daigleetal06} presented an improved sky subtraction scheme by measuring the sky in several places around the galaxy, fitting a polynomial and interpolating the sky over the galaxy to be subtracted from each channel map. This scheme cannot be used for our data set since M\,83 ($\sim 12\arcmin$ major diameter) is larger than the \ghafas\ field of view. Hence, we measured the sky spectrum by observing one cycle of a blank field 10\arcmin\ away from the galaxy after every three cycles on the object. This strategy ensures that we can measure sky variations with time and airmass, and it enables us to statistically measure the sky spectrum. As the sky field is observed in a position slightly offset from the galaxy, the structure (if any) of the sky over the field of view could be different from that over the galaxy. We opt for not making any assumptions about the sky variation over the galaxy and its surroundings, and subtract from the galaxy cube a mean sky spectrum.

We calculate the mean $\overline{s}_{\lambda}$ and the standard deviation $\Delta_{\lambda}$ of each channel in the sky cube. The mean sky profile shows a clear peak at $\sim 6577$ \AA, in agreement with the position of an OH doublet (unresolved in our data) according to \cite{Osterbrocketal96}.

\begin{figure}
     \centering
     \includegraphics[height=5cm,width=8cm]{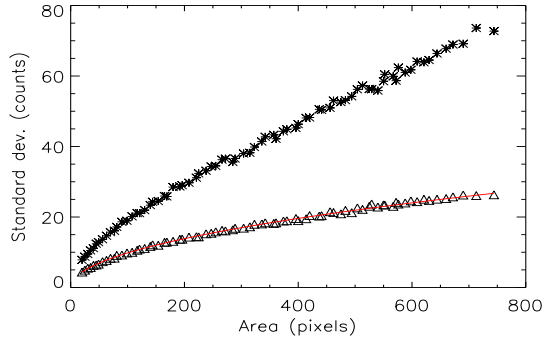}
     \caption{Every point in this figure is the standard deviation of the number of counts in 2000 groups of $n_{pix}$ pixels, measured in two ways. The groups symbolized by triangles are formed by pixels randomly selected across the field of view, and they do follow the behavior predicted by the central limit theorem (solid curve). The asterisks represent groups where pixels have been taken in boxes, so they all belong to the same area of the image.}\label{structure} 
\end{figure}

We then subtract $\overline{s}_{\lambda} \cdot n_{pix}$ to correct the profile of an integrated \HII\ region with $n_{pix}$ pixels, which according to the \emph{central limit theorem}, should have a standard deviation of $\Delta=\Delta_{\lambda} \cdot \sqrt{n_{pix}}$ in the case of a random distribution. Figure~\ref{structure} demonstrates that this is not the case in our data, as it shows the study done for groups of pixels ranging from 20 to 750 in two different ways. The first one is picking random pixels within a channel of the sky cube while the second one picks a box with the same number of pixels. Doing this procedure 2000 times allows us to make statistics about the mean value and the standard deviation of the groups. We see clearly in  fig.~\ref{structure} that the randomly picked pixels follow the expected behavior marked by the central limit theorem, while the boxes have bigger scatter due to different areas having systematically different number of counts, i.e. due to the sky having structure. As we are subtracting the sky to single integrated \HII\ regions, the latter scatter provides us with a reliable measurement of the error committed in subtracting the sky.

\subsection{\ghafas\ instrumental response}
  \label{sec:response}
\begin{figure}
     \centering
     \includegraphics[height=7cm,width=8.4cm]{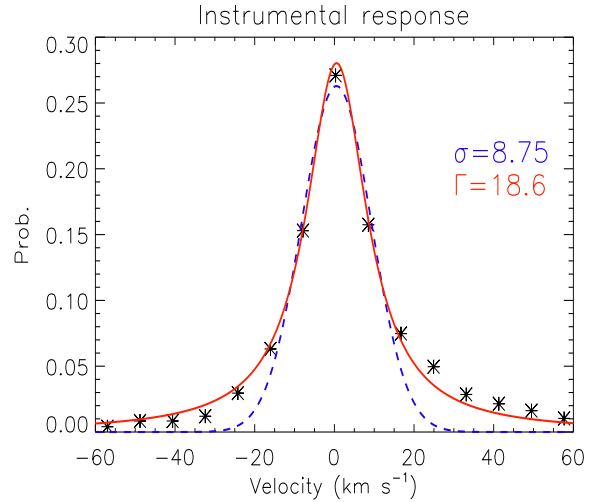}
      \footnotesize{\caption{Instrumental Response Curve of \ghafas\ : asterisks represent the calibration line Ne$\lambda$6568.95 \AA\ , the dashed line is the fit to a Gaussian and the continuous line the Lorentzian fit. Although both lines are correct for the core of the line, the Lorentzian performs better in the wings.}{\label{response}}}
   \end{figure}

We measure the Instrumental Response Curve (IRC) of \ghafas\ by observing the calibration Neon lamp line at 6598.95 \AA. To this line we fit both a Gaussian and a Lorentzian, getting a velocity dispersion for the Gaussian $\sigma=8.75$ \kms\ (equivalent to  FWHM=20.6 \kms\ ) and  FWHM=18.6 \kms\ for the Lorentzian.
As shown in fig.~\ref{response}, both functions fit well the core of the IRC, giving similar results for the FWHM, but in the wings the Gaussian performs clearly worse than the Lorentzian, as has been known for many years \citep[e.g.][]{Bland-Hawthorn95}.

We will demonstrate in sect.~\ref{sec:fitting} that these residual wings are indeed very important for calculating the shape of the profiles.

\subsection{Building a catalogue of \HII\ regions}
\label{sec:leonel}
To detect and catalogue the \HII\ regions in M\,83, we use the software package {\sc REGION} written by Clayton Heller \citep{Rozasetal00}. This package is applied to the reduced, flux calibrated and continuum subtracted \Ha\ image and identifies the regions, measures the position of the center of each region, obtains the area of the region in pixels, and calculates the total \Ha\ flux, integrating over all the pixels in the region and subtracting the sky value. The part of the program that calculates the \HII\ region flux is capable of dealing with an irregular background, so we define 43 circular areas where the program estimates the average value. The background value applied to each \HII\ region is that corresponding to the mean value of the nearest sky area.

\begin{figure}
 \centering
{\includegraphics[width=8.5cm]{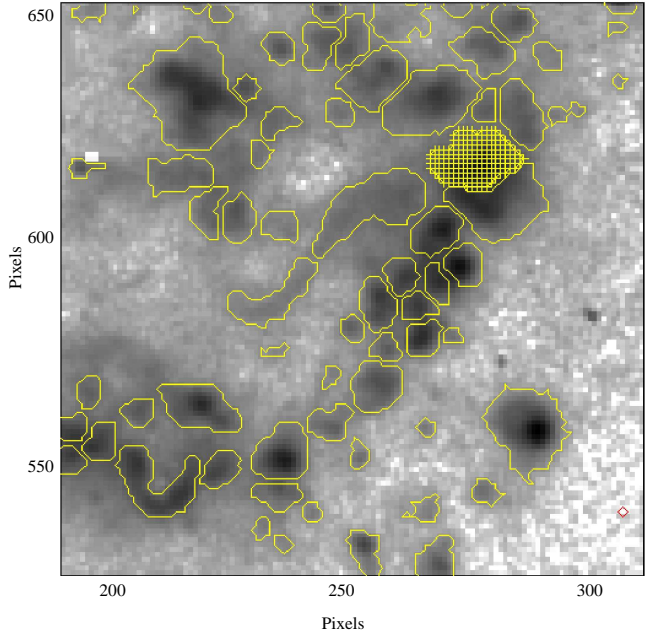}}
\caption{Example of \HII\ regions selected by the program {\sc REGION}. The vast majority of the borders were selected automatically by the program with the parameters determined after several iterations. In a few cases the borders were edited manually especially to separate some overlapped regions.}\label{leonel2}
\end{figure}

The analysis is done in an automatic process, but some fine adjustments to the regions are allowed in an interactive way, intended to modify their size by manual editing. So, with the help of the user, the "apertures" can have an arbitrary shape and not precisely circular, as seen in fig.~\ref{leonel2}.

In order to calculate the boundaries of a region, we calculate the standard deviation in the nearest background and extend the aperture until the surface brightness decreases to a value of twice the standard deviation over the background level. Any pixel with a value lower than $2.5\times10^{-16} erg s^{-1} cm^{-2}$ is considered part of the diffuse ionized gas, hence the flux integration within the aperture is done only with pixels whose signal is higher than this critical level. In addition, all the regions with areas smaller than 20 pixels or integrated flux smaller than $3.5\times10^{-15}  erg s^{-1} cm^{-2}$ are discarded.

In some cases it is difficult to decide whether a feature is the product of a single \HII\ region or the overlap of two or more smaller regions. In those cases, a plot of the image surface is produced. If more than one peak is visible, with a valley deeper than 75\% of the height of the smallest peak, they are separated and treated independently. Otherwise, the reagion is treated as a single region.

Using these criteria 1029 regions are found over the face of the galaxy, of which $\sim 350$ are inside the region covered by the \ghafas\ data cube, i.e. the central 2.3 kpc radius. We cross match the \HII\ region catalogue with the cube image from \ghafas, and find that 193 regions can be identified in the Fabry-Perot cube. We further note that the \ghafas\ observations are not as deep as the \Ha\ image, and some regions have to be omitted from our analysis since they cannot be identified clearly.

\section{Analysis}
\label{sec:analysis}

\subsection{Extraction of line profiles}
\label{sec:profiles}

 Due to the fact that the Fabry-Perot data acquisition is intrinsically different from the narrow-band image, we have restricted our study to those regions which displayed an integrated line profile with amplitude-over-noise $A/N > 10$. The $A/N$ ratio has been measured fitting a single Gaussian to the line profile, been $A/N$ the amplitude of the Gaussian divided by the standard deviation of the spectrum outside $3\sigma $ from the peak.  Although we will discuss in the next section that fitting to a Gaussian is not valid when studying the shape of an integrated line, it is a satisfactory approximation to make a $A/N$ estimation.

\begin{figure*}
 \centering
\includegraphics[width=.99\textwidth]{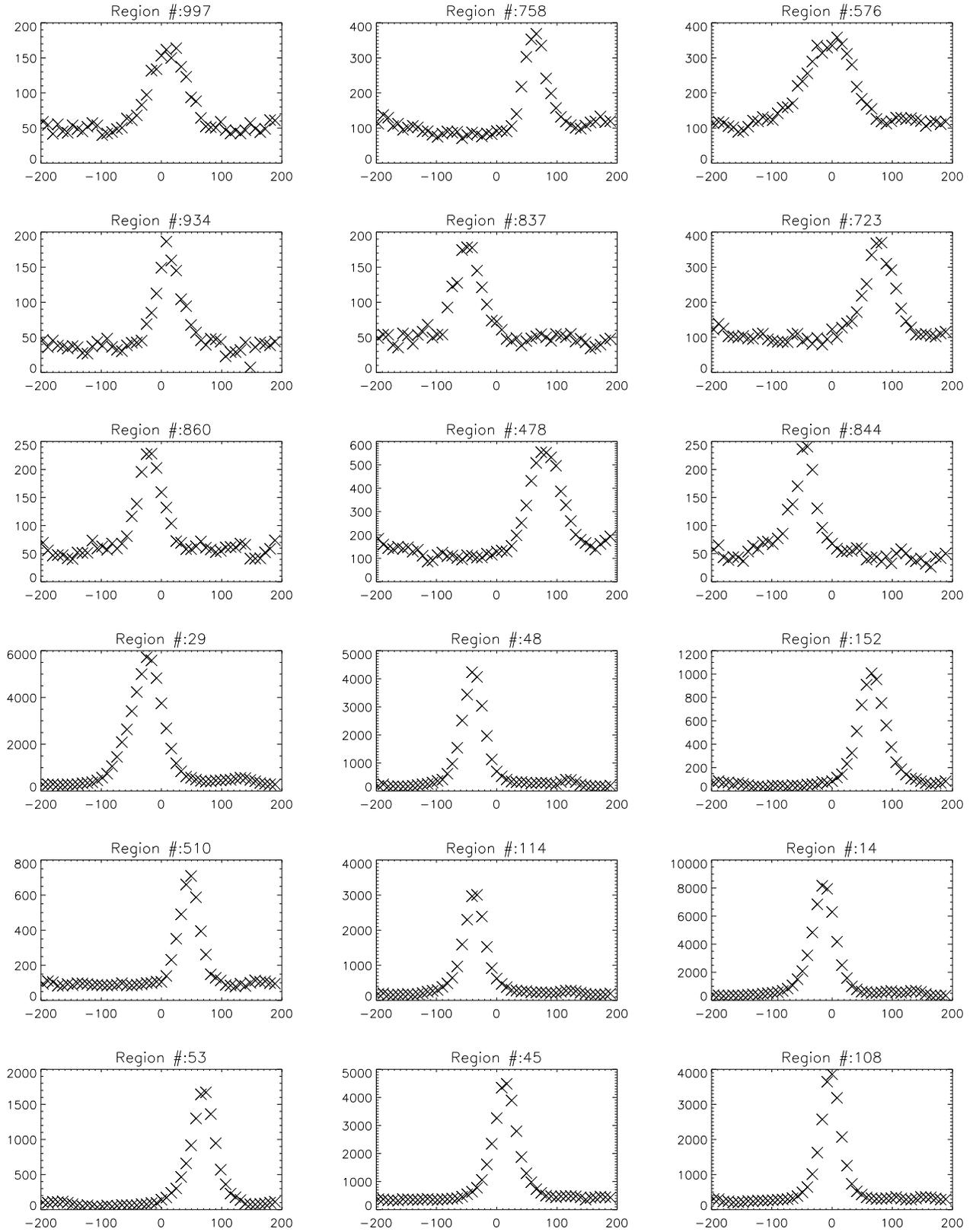}
\caption{Examples of \HII\ region integrated emission line profiles extracted as described in sect.~\ref{sec:profiles}. Top three files are the 9 regions with worst A/N, while bottom three rows show the 9 regions with best A/N.}\label{fig:examples}
\end{figure*}

Applying these two selection criteria leaves us with a final sample of 157 \HII\ regions for which we have good quality \Ha\ emission line measurements. Eighteen of those regions are presented as an example in fig.~\ref{fig:examples}.

As illustrated in fig.~\ref{fig:selected}, the selected regions are distributed over the entire \ghafas\ field, and thus allow us to study the very center of the galaxy, the arms, inter-arms, isolated \HII\ regions as well as those in groups.

\subsection{Fitting procedure}
\label{sec:fitting}

We analyse the physical properties of the \HII\ regions by fitting single and multiple Gaussian components to the integrated \Ha\ emission line profiles.  An observed emission line $I_{\rm obs}$ is formed by the intrinsic emission line profile from the \HII\ region $I_{\rm real}$ convolved with the IRC:
\begin{equation}
\label{convolution}
I_{\rm obs}=I_{\rm real} \otimes {\rm IRC},
\end{equation}
where $I_{real}$ contains the thermal and non-thermal broadening. 

 To study the  non-thermal broadening effect, a correction for the contribution of the instrumental and thermal broadening is needed. It is common to assume that the thermal broadening and the IRC are described by Gaussians with velocity dispersions $\sigma_{\rm th}$ and $\sigma_{\rm inst}$, respectively \citep[e.g.][]{{Rozasetal98},{Relanoetal05},{Martinez-Delgadoetal07}}. In such idealised case, the convolution of Gaussians leads to quadratic co-addition of the velocity dispersions and thus the non-thermal velocity dispersion can be calculated as:

\begin{equation}
\label{quadrature}
\sigma_{\rm nt}^2=\sigma_{\rm obs}^2-\sigma_{\rm inst}^2-\sigma_{\rm th}^2-\sigma_{\rm n}^2,
\end{equation}
where $\sigma_{\rm inst}=8.75$ \kms\ (see sect.~\ref{sec:response}) and, assuming an electron temperature of 7000 K for the \HII\ regions of M\,83 \citep{Bresolinetal05}, $\sigma_{\rm th}=7.63$ \kms\ \citep[for details about the calculation of thermal broadening see ][]{ODellandTownsley88}. The last term, $\sigma_{\rm n}$ corrects for the fine structure of the \Ha\ line, described in detail in \cite{Cleggetal99} and, for very narrow \Ha\ lines, in \cite{CummingandMeikle93}. Simulations based on the seven fine-structure components for $N_e\sim 10^2$ and $T_e\sim 7000$ K yields a value for this correction of $\sigma_{\rm n}=3.2$ \kms\ . \black

The main caveat of the quadratic subtraction is the assumption that the IRC of a Fabry-Perot interferometer is well represented by a Gaussian, when it is better described by a Lorentzian. 

Figure~\ref{response} shows that although a Lorentzian is a better representation than a Gaussian, there are asymmetries in the \ghafas\ IRC. Therefor, we prefer not to assume any analytical function, using instead the measured non parametrised IRC,  averaged over the full field of view. We note furtheremore that the IRC does not vary considerably across the field. Only for comparison reasons, all 157 \HII\ regions were also fitted according to eq.~(\ref{quadrature}), although we stress once again that this is a poor approximation.

The fits are carried out using the IDL procedure MPFIT \citep{Markwardt09}, a robust routine for $\chi^2$ minimization using the Levenberg-Marquardt technique. When the best fit is achieved, the result is four parameters (seven if two Gaussians are fitted): continuum, central velocity(s), height(s) and velocity dispersion(s). Each of these parameters has a formal standard deviation around the optimal value, which allows us to perform a Monte-Carlo simulation in order to estimate the scatter expected around the model in each of the channels.

\section{Results and discussion}
\label{sec:results}
\begin{figure*}
 \centering
\includegraphics[width=.99\textwidth]{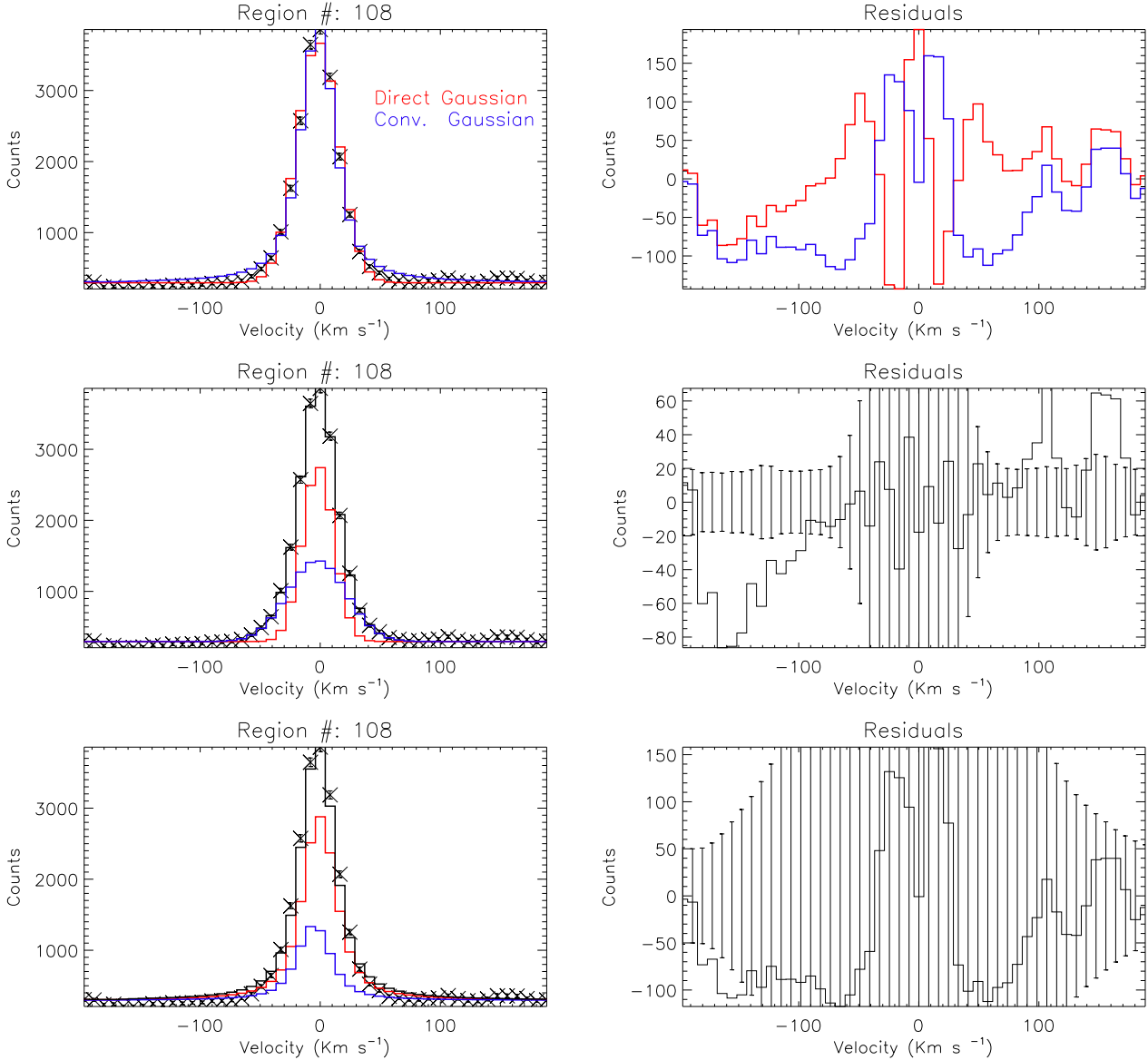}
\caption{Example of the result of the fitting procedure (sect.~\ref{sec:fitting}). \emph{Upper left:} Fit to one Gaussian (blue) and one Gaussian convolved with the instrumental response (red). \emph{Center left:} Fit to two Gaussians, the black line being the addition of both components. \emph{Lower left:} Fit to two Gaussians convolved with the instrumental response. Again the black line is the addition of both components. \emph{Right panels:} Residuals to the fits. Vertical black lines are the expected scatter in each channel due to the addition of three sources: scatter in fitted parameters, sky subtraction and Poisson noise of the incoming photons. }\label{fit_ex}
\end{figure*}

An example of the fit provided by the procedure described in sect.~\ref{sec:fitting} is presented in fig.~\ref{fit_ex} for one and two Gaussians, both using direct Gaussian fitting and the \emph{convolved model}. 

\subsection{Convolution versus quadratic subtraction}
\label{sec:convolution}
The reduced $\chi^2$ of the fits are ranging from 0.8 to 10 for both methods, in 76\% of the regions favorable to the convolution method. The direct Gaussian failed to fit the two most luminous points in 65\% of the profiles due to the weight that the wings have in the fitting, while the convolved Gaussian give a better fit in both the wings and the core. 

On the other hand, reduced $\chi^2$ is up to 10 because our noise is underestimated, as shown by the fact that the dispersion in the continuum is bigger than the individual uncertainties of any of the points. When we use the dispersion of the luminosity of the points in the continuum ($>3\sigma$ away from the center of the profile) as the noise for these points, a $\chi^2$ between 0.7 and 4 is recovered for the direct Gaussian fit while the convolved model reduces to the range 0.7--2. Thus, we can conclude that for the case of one Gaussian, the fit using the convolved method is more robust. 

When performing a two-Gaussian fit to our data, the $\chi^2$ of the fit is improved, but also the uncertainties in the derived parameters are bigger.  The formal uncertainties given by MPFIT  when fitting a single Gaussian are 1--3\% for the height and 2--8\% for the velocity dispersion in the case of single Gaussian fitting, while for the case of the convolved Gaussian the uncertainties are 4--8\% and 2--12\%, respectively. When fitting two Gaussians the uncertainties increase to $\sim 40\%$, even larger when the convolved model is used. This is explained by the fact that the line profile is present, in the best of the cases, in a dozen channels and we are fitting six parameters to it.

When fitting two Gaussians requiring amplitude ratio greater than $1:5$, 32 regions can be fitted with less than $20\%$ error in the velocity dispersion, and when two convolved Gaussians are fitted, 11 regions can be fitted following the same criteria. Thus, we conclude that our observed line profiles are adequately fitted by one single Gaussian only, and inferring an additional Gaussian, will result in very weak parameter determination for both components.

\subsection{Fitted parameters}
\label{sec:parameters}
\begin{figure}
 \centering
\includegraphics[width=8cm, height=8cm]{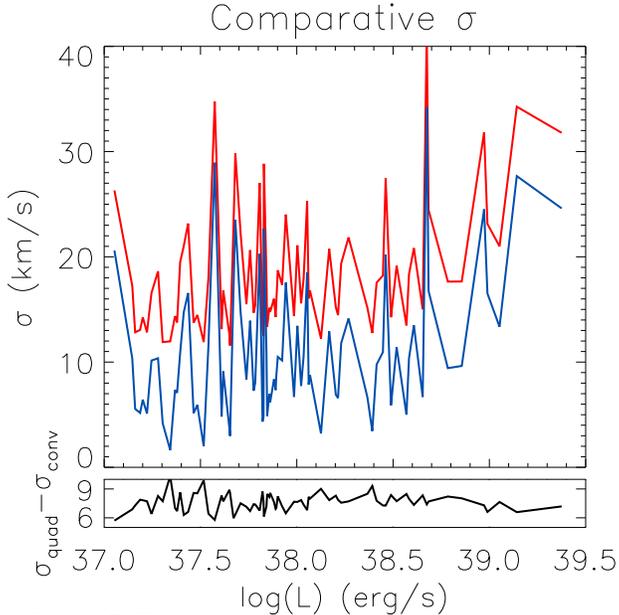}
\caption{The velocity dispersions are displayed against the luminosity for both methods. In red the velocity dispersion calculated by direct fitting of one Gaussian, once corrected from thermal and instrumental broadening. In blue, the velocity dispersion of the same regions calculated with the convolution of a Gaussian with the instrumental response. On the bottom panel the subtraction of both sigmas, giving values around 8 \kms\ for all the regions.}\label{comparision1}
\end{figure}

Comparison of the fitted parameters for both methods show that the position of the center of the line is invariant, an expected result that confirms that in order to get velocity maps it is enough to assume a Gaussian for the line profile. Once applied eq.~\ref{quadrature} to correct the thermal and instrumental broadening from the Gaussian fit, and the thermal broadening from the convolved Gaussian, we obtain the velocity dispersions presented in fig.~\ref{comparision1}. Quadratic subtraction systematically overestimates the velocity dispersion, in full agreement with the recent study presented by \cite{MoiseevandEgorov08}, which predicted a constant overestimation. In the case of \ghafas\, and with the setup used here, the difference is  $\approx 8$ \kms\ (cf., fig.~\ref{comparision1}).

\subsection{$L-\sigma$ relation}
\label{sec:Lsigma}

\begin{table*} 
  \begin{minipage}{160mm}
  \begin{center}
   \caption{Summary of papers relevant for sect.~\ref{sec:Lsigma}. All are based on Fabry-Perot data, with quadratic subtraction of the IRC (details in sect.~\ref{sec:fitting})}\label{table1}
     \begin{tabular}{|c|c|c|c|}
       \hline
       Paper & Sample & Results \\
       \hline
       \hline
       \cite{SmithandWeedman70} & 9 & High velocity dispersions ($\sigma>25$\kms\ )   \\
       \cite{Melnick77} & 16 &  $R\propto \sigma^{2.27}$ \\ 
       \cite{TerlevichandMelnick81}\footnote{Uses data from Melnick (1977)} & 12-15 & $L({\rm H}\beta) \propto \sigma^4$ \\
      
       \cite{Royetal86}\footnote{No mention about the IRC correction} & 47 & $\log(L_{{\rm H}\alpha})= (3.15\pm 0.83) \log \sigma +(35.37\pm 1.17)$ \\ 
       \cite{Hippelein86} & 43 &  $\log(L_{{\rm H}\alpha})= (6.6\pm 0.4) \log \sigma +(31.1\pm 0.3)$ \\
       \cite{Melnicketal87} & 22 &  $\log(L_{{\rm H}\beta})=(4.92\pm 0.30)\log\sigma+(33.24\pm0.31)$ \\
       \cite{Arsenaultetal90}\footnote{Defines an upper envelope for the $L-\sigma$ relation, not a fit} & 127 & $\log(L_{{\rm H}\alpha})= 2.18 \log \sigma +36.5$ \\
       \cite{Rozasetal98}{\Large $^c$} & 200 & $\log(L_{{\rm H}\alpha})= 2.6 \log \sigma +36.15$ \\
       \cite{Relanoetal05}{\Large $^c$} & 373 &  $\log(L_{{\rm H}\alpha})= (2.0\pm0.5) \log \sigma +(36.8\pm0.6)$ \\
       \hline
     \end{tabular}
  \end{center}
  \end{minipage}
\end{table*}

The relation between luminosity and velocity dispersion, can be described by:

\begin{equation}
\label{eq:Lsigma}
\log(L) = a\cdot \log(\sigma) + b
\end{equation}
 
\begin{figure}
 \centering
\includegraphics[width=9cm, height=12cm]{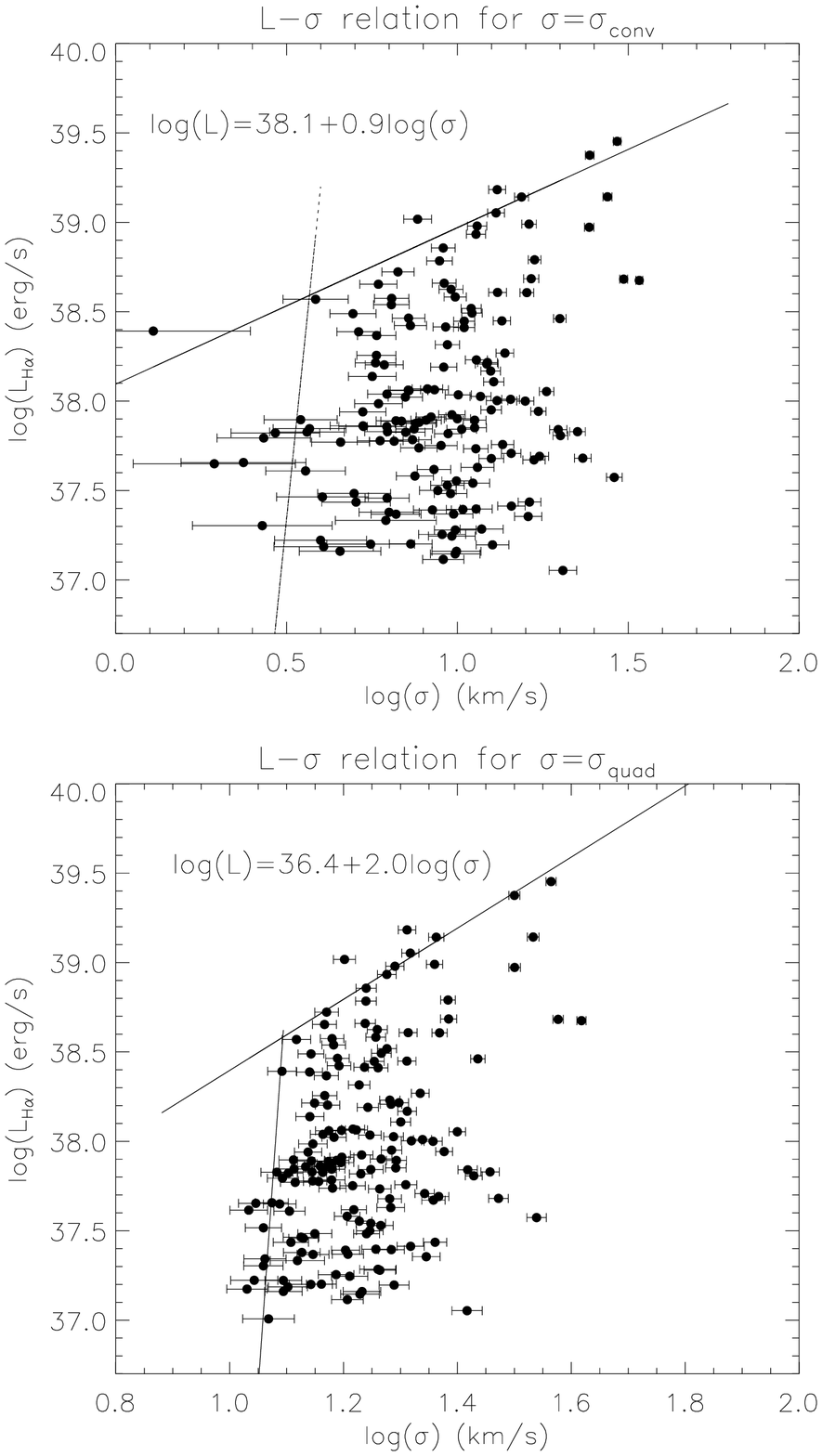}
\caption{The $L-\sigma$ relation for both methods: convolved and direct Gaussian fit. Each point represents one region, the solid line is the fit of the upper luminosity envelope. A clearly different trend is found for the regions with lower velocity dispersion, as showed by the dashed line.  Note the difference in the x-axis for both panels.}\label{fig:Lsigma}
\end{figure}

This relation has been a matter of discussion for the last four decades. Table~\ref{table1} summarizes the main articles related with the $L-\sigma$ relation, the sample sizes and the results they reported. All these studies are based on Fabry-Perot observations, and used the quadratic subtraction to correct from the instrumental response. While the earlier works pointed to a correlation between both variables, with slope between 2 and 7, as observations were refined studies showed that the best linear relation found when plotting $\log(\sigma)$ against $\log(L)$ is that to the upper envelope of the points \citep{{Arsenaultetal90},{Rozasetal98},{Relanoetal05}}. These latter works showed a scatter diagram very similar to fig.~\ref{fig:Lsigma}, where we present the $L-\sigma$ relation for both the convolved and the quadratically subtracted method, assuming for the Luminosity a distance of $4.5$ Mpc \citep{Thimetal03}. \cite{Beckmanetal00} described this envelope as the locus of the regions with minimum velocity dispersion for a given luminosity. Such a minimum velocity dispersion suggested that those \HII\ regions near the envelope are in virial equilibrium, which implies that the great majority of the regions are not in virial equilibrium. \black This is entirely plausible considering the many contributions of dynamically impelled gas to the emission line profiles in general, as we have seen in sect.~\ref{sec:intro}. \black
 
As \cite{Arsenaultetal90}, \cite{Rozasetal98} and \cite{Relanoetal05}, we do find an upper envelope for the $\log(L)-\log(\sigma)$ diagram. Our data do not allow us to discuss whether virial equilibrium can be applied to the regions near the envelope, but only to report that such an envelope, with minimum velocity dispersion for a given Luminosity exists, as claimed by the works cited earlier.\black\  The slope of this envelope depends on the method used to correct the influence of the IRC. For the convolved model we find $\log(L_{{\rm H}\alpha})= (0.9\pm 0.2)\log(\sigma)+(38.1\pm 0.1)$, and we insist on this as the best of the two methods, as it takes into account the full shape of the IRC. On the other hand, and to be able to compare with previous studies, the envelope for the quadratic subtracted $\sigma$ is also calculated, giving $\log(L_{{\rm H}\alpha})= (2.0\pm 0.3)\log(\sigma)+(36.4\pm 0.4)$, in perfect agreement with the latest results in Table~\ref{table1}.

\subsection{\HII\ regions as a possible distance estimator}

Since the earliest observational studies of the $L-\sigma$ relation the idea of using it to help formulate a distance indicator has been discussed by several authors \citep[e.g.][]{{Melnicketal87},{ArsenaultandRoy88}}.

In their article \cite{Beckmanetal00} showed that when histograms of the number of \HII\ regions are plotted vs. luminosity, this luminosity function for a given galaxy has a break at $\log(L)=38.6(\pm 0.1) dex,$ a break first reported by \cite{Kennicuttetal89} and repeatedly confirmed by other authors \citep[e.g.][]{{Rand92},{Rozasetal96}, {Bradleyetal06}}. \cite{Beckmanetal00} put forward the luminosity of this break as a possible standard candle, by analogy with, but more powerful than the standard candle offered by the knee of the luminosity function of planetary nebulae. All the above mentioned studies are based on reasonably face-on galaxies, ( $i < 50^\circ$), in order to optimize the number of regions detected, prevent crowding effects, and minimize the extinction by dust in the disks of the galaxies. These effects will limit, to some extent, the utility of the 38.6 dex break, notably for galaxies with higher inclinations, but the validity of the result for low inclination galaxies appears solid.

The data displayed in fig.~\ref{fig:Lsigma} show that the linear upper envelope referred to above sets in at log(L) between 38.5 and 38.6. This value is based on a distance to M\,83 based on measured Cepheid variables. This is, to our knowledge, the first study that shows a break point in the envelope of the $\log(L)-\log(\sigma)$ diagram. Without venturing into the underlying physical cause of this change in slope its existence suggests that there is such a cause, and tends to reinforce the use of the $\log(L)-\log(N)$ histogram, or indeed the envelope of the $\log(L)-\log(\sigma)$ itself, with their respective break points, as possible versions of a secondary standard candle.

The uncertainty in the determination of the point where the change of regime happens in our data yields errors of the order of 0.4 Mpc, comparable with the 0.2 Mpc error of the Cepheid measurement by \cite{Thimetal03}. A better determination of this point for M\,83 should be possible, given the fact that our $3.4'\times 3.4'$ \ghafas\ field of view covers only $\sim 8\%$ of the disk in M\,83.

\subsection{\Ha\ equivalent widths}
Following the formalism developed in the Starburst99 (SB99) model \citep{Leithereretal99} some crude assumptions can be used to obtain a rough idea about the ages of the observed HII regions from our data. We acknowledge that the situation is very complex not only due to observational difficulties but also due to dust scattering effects on the continuum levels, nevertheless, as a mere exercise we can assume comparable dust content in the spiral arms and the bar, an instantaneous burst of star formation for the HII regions, and a constant average metallicity, Z = 0.04 \citep{BresolinandKennicutt02} and use the measured \Ha\ equivalent widths to estimate integrated average ages of \HII\ regions in M\,83. Comparing the SB99 models with a typical value of the EW of 60 \AA\ for those HII regions residing in the spiral arms, we obtain a mean age of 9 Myr. A similar exercise for the \HII\ regions in the bar yields an average EW of 20\AA\ , which corresponds to a mean age of 12 Myr. Although these values are lower limits, the systematic difference between the EWs in the arm and bar region suggests that the HII regions in the spiral arms are typically younger than those in the bar of M\,83. Leakage from other orders of interference, due to imperfectly efficient interference, affect both our measured \Ha\ continuum luminosity, resulting in an underestimation of the EWs since the equivalent width (the ratio of the \Ha\ luminosity to that of the nearby continuum) is very sensitive to the continuum value, Fabry-Perot observations should produce values for the EW which are systematically low, so the actual ages should be interpreted with caution.

\black

\begin{figure*}
  \centering
  \subfigure
  {
    \includegraphics[scale=0.85]{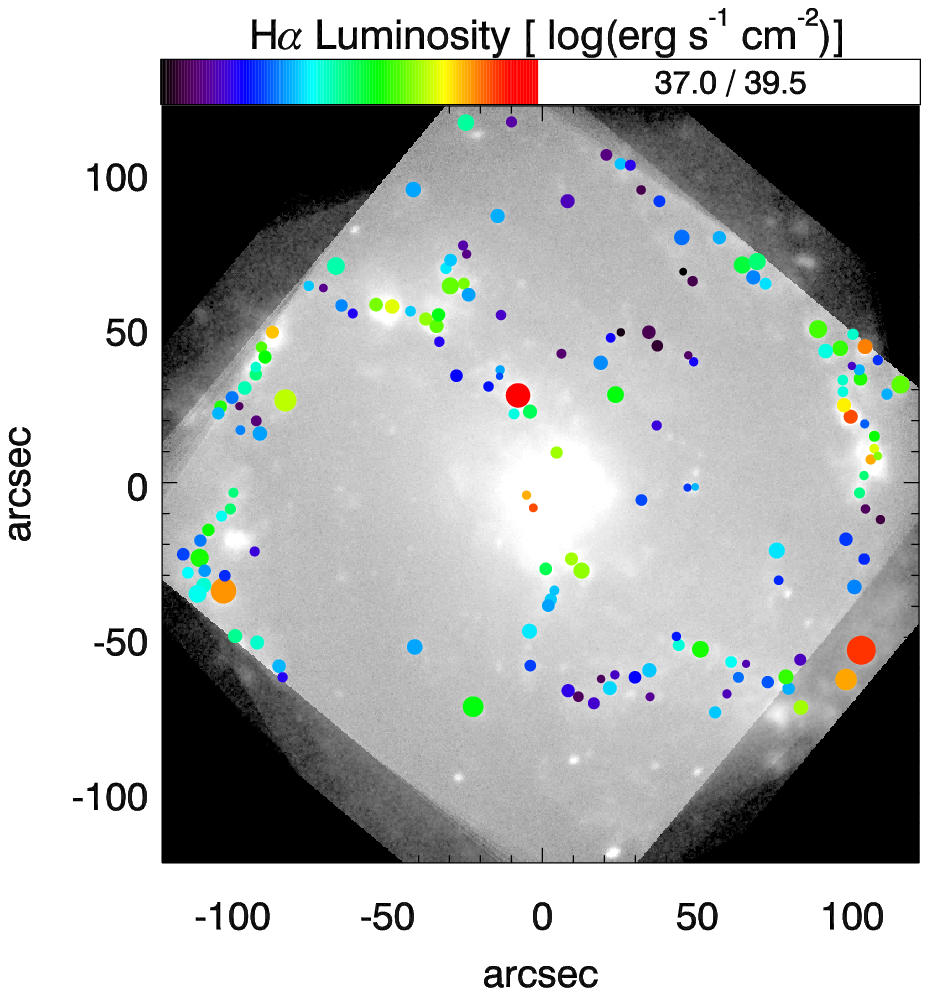}
  }

  \subfigure
  {
    \includegraphics[scale=0.85]{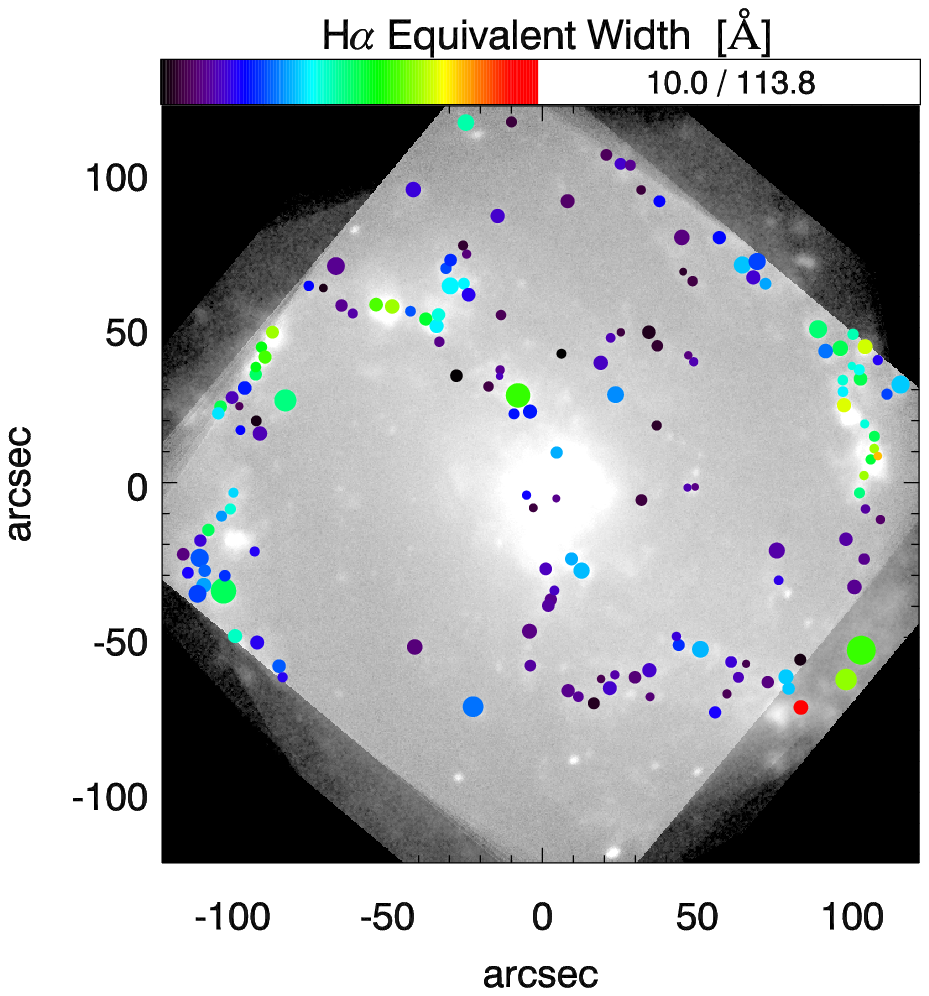}
  } 

  \caption{\Ha\ luminosity and Equivalent width map of all the 157 regions within the \ghafas\  field of view. }\label{other_dependencies_1}
\end{figure*}

\section{Conclusions}
\label{sec:conclusions}
We have presented an improved method for analysing emission line profiles from Fabry-Perot interferometers, and have applied it to \ghafas\ Fabry-Perot data for the central 2.3 kpc radius of the nearby Sc galaxy M\,83. 
We have considered the true instrumental profile without any parametrisation, and derived observed emission line profile parameters by a convolution of a Gaussian with the instrumental response.

From a narrow-band \Ha\ image, a catalog of \HII\ regions was extracted and we were able to determine the flux, position, and area for 1029 star forming \HII\ regions. Improving some of the Fabry-Perot data reduction procedures (sect.~\ref{sec:reduction}) we have measured accurate integrated \Ha\ emission line profiles for 157 \HII\ regions in the \ghafas\ field. 

Convolving with the IRC before fitting is the more realistic approach and clearly gives better fitting results in terms of smaller $\chi^2$ for 76\% of the regions. When fitting two Gaussians, the errors in the determination of the line parameters increase to more than 40\%, due to low number of degrees of freedom, and for this reason we ruled out successful multiple-Gaussian fitting. Following the commonly used procedure, when fitting Gaussians in a direct way, the IRC was assumed a Gaussian and quadratically subtracted. The direct fitting method leads to an overestimation of $\sigma$ by a constant value of $\sim 8$ \kms\ 
. We used the results from our improved line fitting approach to study the $L-\sigma$ relation and found no significant linear correlation between these two variables, but an upper envelope defined by $\log(L_{{\rm H}\alpha})=1.3 \cdot \log(\sigma)+37.6$. Our data is fully consistent with the previous estimations of \cite{Arsenaultetal90}, \cite{Rozasetal98} and \cite{Relanoetal05}, when applying direct Gaussian fitting.

The upper envelope emerges in the $\log(L)-\log(\sigma)$ diagram at luminosities higher than $10^{38.5}erg\, s^{-1}$. 

\cite{Beckmanetal00} suggested that this luminosity might correspond to the transition from an ionization bounded regime to a density bounded regime (i.e. from optically thick to optically thin). We now understand, in part from work by \cite{Giammancoetal04} that the simple Str\"omgren  sphere scenario on which this concept is based is too simple, and that the strongly inhomogeneous structure of \HII\ regions must be taken into account when explaining any product of radiative processes within them. Nevertheless an explanation in which there is an increase in the escape fraction of ionizing photons from \HII\ regions above this luminosity is still a reasonable way to account for a variety of observations, even though more sophisticated modelling will be needed to give final quantitative answers.Our sample has extended previous \HII\ region analyses, by including a significant number of faint \HII\ regions with their corresponding values of $\sigma$, and this has allowed us to better pin down the luminosity at which a clean upper envelope in the $L-\sigma$ relation emerges. Using the measured value as a secondary standard candle we find that the uncertainty in the position of the break point yields a contribution to the  uncertainty in the distance to M\,83 (0.4 Mpc) which is not much greater than the uncertainty in the Cepheid distance itself (0.2 Mpc). As our observations cover less than 10\% of the optical disk of M\,83 this suggests that a more complete data set could be used to reduce this uncertainty.

\section*{Acknowledgments}
We thank the anonymous referee for insightful comments that improve the quality of our paper.

We thank the ING staff for all the support we received during our observations, in particular Juerg Rey, Andrew Cardwell, Diego Cano, and Miguel Santander Garcia. J.B-H. acknowledges financial support from NOTSA. K.F. acknowledges support from the Swedish Research Council (Vetenskapsr\aa det). J.E.B. acknowledges support from the Spanish Ministry of Education and Science Project AYA2007-67625-CO2-01 and the IAC project P3/86. C.C. and O.H. acknowledge support from NSERC, Canada and FQRNT, Qu\'ebec. 

\bibliographystyle{astron}
\bibliography{bibliografia}

\label{lastpage}
\end{document}